\documentclass[preprint]{article}
\usepackage{graphicx}
\usepackage{dcolumn}
\usepackage{bm}
\usepackage{amsmath}
\usepackage{amssymb}
\usepackage{multirow}
\usepackage{caption}
\usepackage{epstopdf}
\usepackage{hyperref}

\begin{document}
{\setlength{\oddsidemargin}{1.2in}
\setlength{\evensidemargin}{1.2in} } \baselineskip 0.55cm
\begin{center}
{\LARGE {\bf Cylindrical gravastars with Kuchowicz metric potential }}
\end{center}
\date{\today}
\begin{center}
  Meghanil Sinha*, S. Surendra Singh \\
Department of Mathematics, National Institute of Technology Manipur,\\
Imphal-795004,India\\
Email:{ meghanil1729@gmail.com, ssuren.mu@gmail.com}\\
 \end{center}

 \textbf{Abstract}: Mazur and Mottola's gravastar model represents one of the few serious alternatives to the traditional understanding of the black hole. The gravastar is typically regarded as a theoretical alternative for the black hole. This article investiagtes the creation of gravastar(gravitational vacuum star) within the realm of cylindrically symmetric space-time utilizing the Kuchowicz metric potential. A stable gravastar comprises of three distinct regions, starting with an interior region marked by positive energy density and negative pressure $(p=-\rho)$ which is followed by an intermediate thin shell, where the interior negative pressure induces a outward repulsive force at each point on the shell. Ultra-relativistic stiff fluid makes up the thin shell governed by the equation of state(EoS) $(p=\rho)$, which meets the Zel'dovich criteria. And then comes the region exterior to it which is total vacuum. In this scenario, the central singularity is eliminated and the event horizon is effectively substituted by the thin bounding shell. Employing the Kuchowicz metric potential we have derived the remaining metric functions for the interior region and the shell regions yielding a non-singular solution for both the regions. Additionally, we have investigated various characteristics of this shell region including its proper shell length, the energy content and entropy. This theoretical model successfully resolves the singularity issue inherent to the black holes. Therefore, this gravastar model presents a viable alternative to the traditional black holes, reconciled within the context of Einstein's theory of General Relativity.\\
 
 
 \section{Introduction}\label{sec1}
 
  Study on the ultimate fate of the compact stars holds considerable significance within the domains of cosmology and astrophysical research. According to the Virial theorem, a star can maintain stability if its gravitational potential energy is equal in magnitude to minus twice its thermal energy. After depleting its fuel, a star faces significant challenge in counter-balancing the inward pull of its own gravity. Gravitational collapse gives rise to the formation of various celestial objects, which include the stars, the white dwarfs, and the black holes. Within the framework of General Relativity(GR), black holes have emerged as the most fascinating compact objects in cosmology, with their formation being directly contingent upon the mass of the progenitor collapsing star. Due to its incredibly high density and extremely compact nature, nothing not even light can be able to escape the intense gravitational pull of a black hole once it falls within its boundary. The inevitable presence of a singularity at the end-point of gravitational collapse has long been a source of discomfort and theoretical challenge for the astrophysical community. Thus Mazur and Mottola introduced a novel theoretical model to overcome the singularity problem, where conventional physical laws cease to apply that arises in the penultimate phase of a star's collapse, the concept of gravastar(gravitational vacuum star) \cite{Mazur,Mottola}. It opened up a new arena in gravitational physics.\\
 A gravastar resembles a black hole in many ways, but lacks an event horizon, and thereby avoiding the associated singularity problem. In the gravastar scenario, quantum vacuum fluctuations are anticipated to significantly influence the physics of gravitational collapse. It is believed that a phase transition takes place, resulting in a core exhibiting repulsive properties that counteracts the collapse, thereby preventing the creation of the point of no return and the singularity \cite{Event,Singularity}. The phase transition is predicted to occur extremely near the threshold of $\frac{2m}{r}=1$, making it virtually indistinguishable from a genuine black hole to an external observer. The paradoxes surrounding singularities and event horizons have longstanding concerns among the scientists since the discovery of black holes. Thus gravastar has emerged as a particularly fascinating compact object for modern research, as it offers a potential solution to the longstanding issues of singularities and event horizons.\\
 Within the gravastar's interior, region of negative pressure generates a repulsive force that effectively halts the formation of a singularity. A thin shell, composed of ultra-relativistic matter, marks the boundary at the end of the gravastar's internal region. Zel'dovich initially introduced this concept of stiff fluid to describe high-pressure matter within the cosmological framework of a cold, baryonic Universe \cite{Zel1,Zel2} and then the exterior region of the gravastar is completely devoid of matter, being a perfect vacuum. Following Mazur and Mottola's proposal of a five-layered gravastar model, Visser and Wiltshire simplified it to a three-layer model and demonstrated its dynamical stability against gravitational perturbations \cite{Visser}. We can also see the generalizations of the stability of the gravastar model against different exteriors \cite{Carter}. The stability of gravastars was also analyzed by examining their response to axial perturbations. Additionally, the exploration of the linearized stability analysis of gravastars within the non-commutative geometry frameworks has also been done\cite{Lobo}. We can see the construction of gravastars within the $(2+1)$ dimensional GR framework \cite{FR}. A charged gravastar model was proposed, allowing for the conformal motion, with its exterior space-time being  characterized by the Reissner-Nordstr\"{o}m solution \cite{Us}. The properties of charged $(2+1)$ dimensional gravastars in anti-de Sitter space have also been investigated\cite{Islam}. In the context of Einstein's General Relativity, gravastars have been successfully explored in the higher dimensional space-time. The properties of neutral gravastars in higher dimensional space-time have been examined \cite{Ray}. The characteristics of charged gravastars admitting conformal motion have also been investigated in higher dimensional space-times \cite{Bhar}. Additionally, charged gravastars in higher dimensions have also been explored without the assumption of the conformal motion \cite{Ghosh}.\\
The gravastar model represents comparatively fewer theoretical difficulties and also meets the stability requirements necessary for its existence. The next step was to explore he observational evidence that could support or confirm the existence of gravastars. One potential approach for detecting gravastars is through the use of gravitational lensing techniques \cite{Kubo}. It was discovered that microlensing effects of a gravastar could produce a maximal luminosity significantly greater than that black hole having the same mass \cite{Kubo}. Two detection models for gravastars were proposed, with model 1 involving the calculation of visual representation of the companion object orbitting the gravastar, which depend on whether it might be having unstable photon orbits or not, considering optical transparency of the thin shell surface. Whereas model 2 involves calculating the microlensing effects, specifically the changes in total luminosity. The model reveals that the maximal luminosity can be significantly greater than that of a black hole of equivalent mass. Recently, the LIGO interferometric detectors captures the ring-down signal form the gravitational wave event GW150914 \cite{BP}. Notably, a striking prediction have been made with high possibility that objets lacking an event horizon, such as gravastars, are likely the sources of these detected gravitational waves \cite{Card}. However due to our limited understanding of the perturbative analysis of rotating gravastars, we cannot conclusively attribute the source of the detected gravitational waves to merging gravastars \cite{Chir}.\\
 Typically, researchers utilize the static and spherically symmetric interior and exterior Schwarzschild configurations to the Einstein field equations in their analyses. In contrast, cylindrically symmetric static solutions that incorporate both translational and rotational symmetries along and about the axis are relatively less explored within the GR community. The static, spherically symmetric vacuum solutions were discovered by Weyl and Levi-Civita early in the last century \cite{Weyl,Levi}. Their work focussed on the broader issue of static, axially symmetric geometries. Our goal is to develop a gravastar model that incorporates a varying cosmological constant within the realm of cylindrical framework, with a focus on obtaining solutions that are free from singularities. Recent findings in modern cosmology, based on observational data and theoretical results, suggest that the cosmological constant would provide a viable scientific explanation for dark energy. According to data from the Wilkinson Microwave Anisotropy Probe(WMAP), approximately three-quarters of the Universe's total mass-energy density is comprised of dark energy \cite{Per,Pr,P,Riess}. The dominant theory for dark energy is rooted in the cosmological constant concept, putforth by Einstein in the year of 1917. Zeldovich later interpreted the cosmological constant as a manifestation of vacuum energy arising from quantum fluctuations, with a magnitude of order $\backsim 3\times10^{-56} cm^{-2}$ \cite{YB,Ratra}.\\
 Observations of type Ia supernovae and the other cosmic observations suggest that the Universe's expansion is accelerating, leading to speculation that this phenomenon may be attributed to a positive, non-zero cosmological constant \cite{Cylinder}. The accelerating expansion of our Universe, as indicated by recent observations, has sparked renewed interest in investigating astrophysical objects within the framework of a non-zero cosmological constant. The impact of the cosmological constant on gravastar formation in higher-dimensional space-time, which offers a potential alternative Schwarzschild-Tangherlini black holes have also been investigated in D-dimensional space-times \cite{BK,Tan}. Motivated by this concept, we propose cylindrically symmetric model that incorporates the cosmological constant for the gravastar.\\
 This paper is structured as follows: Section (\ref{sec1}) presents a concise overview of the gravastars along with cylindrical symmetric space-time. Section (\ref{sec2}) outlines the essential mathematical framework for our analysis. Section (\ref{sec3}) examines the interior space-time of the gravastar, while section (\ref{sec4}) and (\ref{sec5}) focus on the shell region and the exterior space-time respectively. Section (\ref{sec6}) addresses the junction conditions. Section (\ref{sec7}) analyzes various key characteristic features of the gravastar within the shell region. Section (\ref{sec8}) deals with the stability analysis of the gravastar model through the study of the surface red-shift and finally the paper concludes with a recap of the main results in section (\ref{sec9}).\\
 
 \section{Mathematical formalism in cylindrical spacetime}\label{sec2}
 
  For solving the Einstein field equations in a cylindrically symmetric space-time, we first need to specify the line element that will serve as the foundation for our analysis. The space-time has a central axis of symmetry, with $z$ representing the co-ordinate for static configurations of the Einstein field equations. Thus the space-time for a general, static and cylindrically symmetric solution is described by \cite{Brito},\\

 \begin{equation}\label{1}
  ds^{2}=-e^{a(r)}dt^{2} + e^{b(r)}dr^{2} + e^{c(r)}dz^{2} + e^{d(r)}d\theta^{2}
  \end{equation}\\
  where $a,b,c$ and $d$ are unknowns. Similarly for spherically symmetric space-time, we define the radial co-ordinate $r$ such that the co-efficient of $d\theta^{2}$ is $r^{2}$ \cite{Cylinder}. This transformation is known as the tangential guage. As a result, by replacing $e^{d(r)}=r^{2}$ the metric can be expressed as, 
  \begin{equation}\label{2}
 ds^{2}=-e^{a(r)}dt^{2} + e^{b(r)}dr^{2} + e^{c(r)}dz^{2} + r^{2}d\theta^{2}. 
 \end{equation}\\
 The coefficients of $dr^{2}$ and $dz^{2}$ being of equal dimensions, we set $e^{b(r)}=e^{c(r)}$ \cite{Cylinder}. Thus by simplifying the metric, we get
 \begin{equation}\label{3}
 ds^{2}=-e^{a(r)}dt^{2} + e^{b(r)}(dr^{2} + dz^{2}) + r^{2}d\theta^{2}.
 \end{equation}\\
 We assume the stress-energy tensor inside the gravastar corresponds to an isotropic fluid, given by
 \begin{equation}\label{4}
 T_{mn}=(\rho+p)u_{m}u_{n}+pg_{mn}
 \end{equation}\\
 where $u_{m}$ denotes the four-velocity component subject to $u_{m}u^{m}=-1$. Here we assume that the cosmological constant has a dependence on the radial distance, i.e. $\Lambda=\Lambda(r)$. As a result, we can have the Einstein field equations as,
 \begin{equation}\label{5}
 G_{mn}=R_{mn}-\frac{1}{2}g_{mn}R + g_{mn}\Lambda = \frac{8 \pi G}{c^{4}}T_{mn}
 \end{equation}\\
 where $R=g_{mn}R^{mn}$ is the Ricci scalar, $G$ represents the Universal gravitational constant and $c$ denotes the speed of light in vacuum. Now, using $G=c=1$ in the relativistic units, from the metric in equation(\ref{3}), we obtain\\
 \begin{equation}\label{6}
 8\pi\rho + \Lambda = -\frac{1}{2}e^{-b}b''
 \end{equation}\\
 \begin{equation}\label{7}
 8\pi p - \Lambda = \frac{e^{-b}}{4r}(a'^{2}r + 2ra'' + 4a')
 \end{equation}\\
 \begin{equation}\label{8}
 8\pi - \Lambda = \frac{e^{-b}}{4}(a'{2} + 2b'' + 2a'')
 \end{equation}
 where prime indicates the differentiation with respect to the radial co-ordinate $r$. We assume the metric potential $e^{a(r)}$ takes the form of the Kuchowicz metric, specifically \cite{Ku},
 \begin{equation}\label{9}
 e^{a(r)} = e^{\alpha r^{2} + 2 \ln \beta}
 \end{equation}\\
 where $\alpha$ and $\beta$ are the arbitrary constants that characterize the Kuchowicz metric. The constant $\alpha$ possesses a dimension of $L^{-2}$(length inverse squared), while $\beta$ is a dimensionless parameter. This metric potential is non-singular and well-defined. Inserting the previous result into equations \ref{6}-\ref{8}, yields the following modified equations\\
 \begin{equation}\label{10}
 8\pi\rho + \Lambda = -\frac{1}{2}(\frac{b''}{e^{b}})
 \end{equation}\\
 \begin{equation}\label{11}
 8\pi p - \Lambda = \frac{1}{4e^{b}}(4\alpha^{2}r^{2} + 12\alpha)
 \end{equation}\\
 \begin{equation}\label{12}
 8\pi p - \Lambda = \frac{1}{4e^{b}}(4\alpha^{2}r^{2} + 2b'' + 4\alpha).
 \end{equation}\\
  The energy-momentum conservation equation takes the form as:
  \begin{equation}\label{13}
  \frac{1}{2}(\rho + p)a' + p' = 0.
  \end{equation}\\

 \section{Interior geometry of gravastar}\label{sec3}
 
  The following section provides a discussion of the internal structure of the gravastar. As previously discussed, the gravastar's structure is made up of three distinct regions: the interior region, the transitional thin shell and the outer region. The gravastar's structure is determined by the parameters of the EoS. We have denoted $r_{1}$ as the radius of the inner region and $r_{2}$ as the radius of the outer region respectively, with an EoS of $p=-\rho$ representing a negative pressure in the interior and the completely vacuum exterior. While in the thin shell, where $r_{1}< r < r_{2}$ denotes the radius and the EoS there is given by $p=\rho$. A repulsive pressure characterized by negative pressure emanates outward radially from the center, counteracting the inward gravitational pull exerted by the shell. The repulsive force present within the interior region is attributed to dark energy candidates \cite{P}\cite{Riess}\cite{Muk}\cite{PP}. By applying this EoS, from equation(\ref{13}) we get\\
 \begin{equation}\label{14}
 p = -\rho = -\rho_{k}
 \end{equation}\\
 where $\rho_{k}$ is the constant representing the central density of the gravastar. As a result, uniform pressure and matter density are maintained throughout the inner space of the gravastar. Utilizing this EoS, from the equations (\ref{10}) and (\ref{11}), we derive the another metric function $e^{b}$ as
 \begin{equation}\label{15}
 e^{b} = He^{\frac{2\alpha}{3}(\frac{9r^{2}}{2}+\frac{\alpha r^{4}}{4}) + rC_{1}}
 \end{equation}\\
 \begin{figure}[h!]
\centering
\includegraphics[scale=0.5]{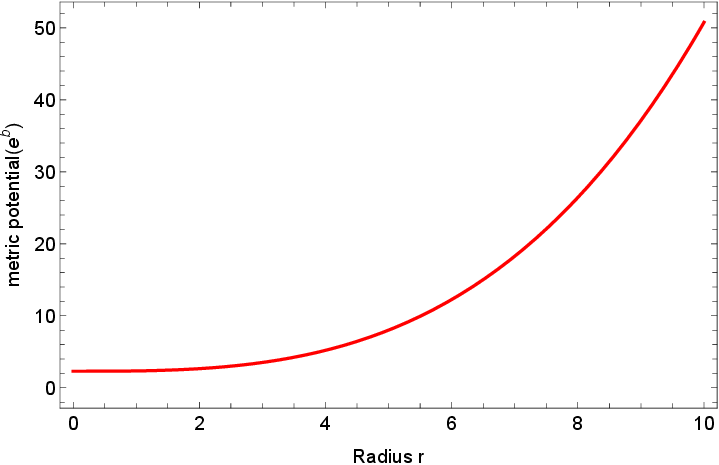}
\caption{The metric co-efficient $e^{b(r)}$ with respect to the radial distance $r$(km) in the gravastar is plotted for $H=10$, $C=1$ and $\alpha = 0.01491932683$ \cite{K}}
\end{figure}
 where $H$ and $C_{1}$ are constants of integration. Evidently, this solution is devoid of singularities. Variation of $e^{b}$ is illustrated graphically in fig (\ref{1}), the increasing behavior of the metric potential in the interior is apparent and is devoid of singularity and also remains finite and positive. At the boundary, for this inner region we set $r=r_{1}$. We can determine the active gravitational mass within the interior region which can be determined through the following expression as,\\
 \begin{equation}\label{16}
 M_{active} = \int_{0}^{r_{1}} 4\pi r^{2} \rho dr = \frac{4}{3}\pi r_{1}^{3} \rho_{k}.
 \end{equation}\\
 
 \section{Intermediate thin shell of the gravastar}\label{sec4}
 
 The gravastar's interior is enveloped by a thin, finite shell comprising ultra-relativistic fluid or soft quanta, which adheres to the EoS,
 \begin{equation}\label{17}
 p=\rho.
\end{equation}\\
This model of the stiff fluid originally employed by Zel'dovich to describe a cold, baryonic Universe \cite{Zel2}, has also been utilized by numerous cosmologists and astrophysicists \cite{BJ,Pra,Mal,TM,Buch}. Throughout the shell(non-vacuum) region, finding a general solution to the field equations poses significant challenges. Thus to obtain a feasible solution, we will employ suitable approximations. Here we assume $0< e^{-b} \ll 1$ in the ultra-relativistic thin shell, where the two space-times merge. Physically, when two space-times join, they must do through a thin, intermediate shell \cite{Is}. In the thin shell, as $r$ approaches 0, any $r$-dependent parameter can be considered $\ll 1$. We obtain the following result from the equations (\ref{10}) and (\ref{11}), using equation (\ref{17}) and the aforementioned approximation as,
\begin{equation}\label{18}
e^{b(r)}=\frac{\alpha^{2}r^{2} + 3\alpha}{2\Lambda}.
\end{equation}\\
\begin{figure}[h!]
\centering
\includegraphics[scale=0.5]{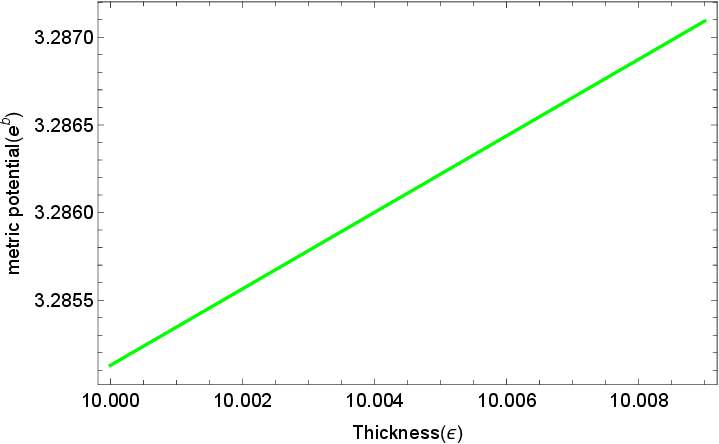}
\caption{Graphical representation of metric coefficient $e^{b(r)}$ in the shell with respect to the thickness parameter $\epsilon$(km) in the gravastar for $\alpha = 0.01491932683$\cite{K}}
\end{figure}
Fig (\ref{2}) pictorially represents the increasing behavior of the metric potential $e^{b(r)}$ in the shell region of the gravastar. It also shows a positive nature there. Again by substituting $p=\rho$ and the Kuchowicz metric potential from equation (\ref{9}) into equation (\ref{13}), we derive the pressure and matter density expressions for the thin shell as
\begin{equation}\label{19}
p = \rho = Ae^{-\alpha r^{2}}
\end{equation}\\
\begin{figure}[h!]
\centering
\includegraphics[scale=0.5]{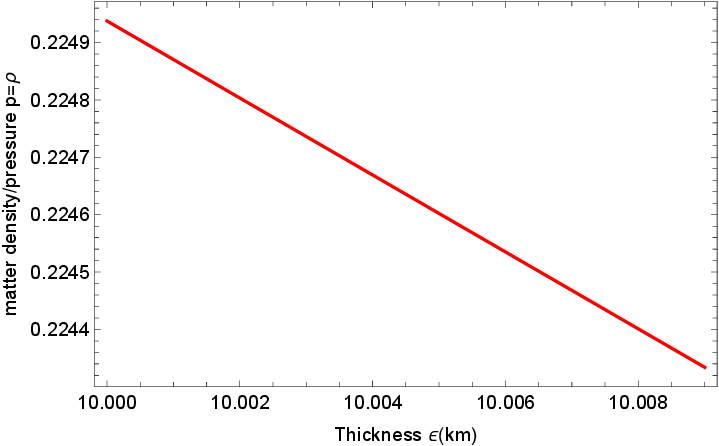}
\caption{Variation of the pressure or matter density with respect to the thickness parameter $\epsilon$(km) in the shell is plotted for $\alpha = 0.01491932683$ \cite{K}}
\end{figure}
where $A$ represents an integration constant. Figure (\ref{3}). illustrates pressure(or matter density) fluctuations across the thin shell. The figure reveals that both the pressure and density are positive everywhere in the shell, exhibiting a steep decline with the increasing radial distance \cite{K}.\\

\section{Exterior geometry of the gravastar}\label{sec5}

The exterior region is characterized by a static, vacuum space-time. The metric for the cylindrical vacuum exterior is expressed as \cite{Lemo},
\begin{equation}\label{20}
ds^{2} = -(\Lambda r^{2} - \frac{4M}{r})dt^{2} + (\Lambda r^{2} - \frac{4M}{r})^{-1}dr^{2} + \Lambda r^{2}dz^{2} + r^{2}d\theta^{2}
\end{equation}\\
where $\Lambda < 0$, denotes the cosmological constant.\\

\section{Junction conditions}\label{sec6}

Gravastar is composed of three distinct regions: an interior region, an exterior and a thin shell in between. The shell serves as the interface or boundary layer which link the two regions of the gravastar. Thus, the shell plays a vital role in maintaining the structural integrity of the gravastar. This configuration forms a geodesically complete space-time, featuring a material shell at the surface that encloses the gravastar. Thus the entire gravastar configuration is described by a single manifold \cite{K}. At the junction, the fundamental junction condition requires a smooth transition between the regions I and III. Although the metric co-efficients are continuous, their derivatives might or might not be continuous \cite{Stab}. We will now apply the Darmois-Israel junction condition to determine the surface stress-energy tensor at the junction \cite{Is,G}. According to the Lanczos equation, the intrinsic surface stress-energy tensor $S_{pq}$ is explicitly determined as follows \cite{L,N,GP,PK},
\begin{equation}\label{21}
S_{q}^{p} = \frac{1}{8\pi}(K_{q}^{p} - \delta_{q}^{p}K_{\kappa}^{\kappa}).
\end{equation}
 The discontinuity in the second fundamental form at the junction can be expressed as \cite{Cylinder2}
 \begin{equation}\label{22}
 K_{\alpha\beta} = K_{\alpha\beta}^{+} - K_{\alpha\beta}^{-}
 \end{equation}\\
 where $``+"$ denotes the exterior and $``-"$ the interior surfaces, whereas the second fundamental form or the extrinsic curvature is defined as,
 \begin{equation}\label{23}
 K_{\alpha\beta}^{\pm} = -n_{\gamma}^{\pm}\Big[\frac{\partial^{2}X_{\gamma}}{\partial \Sigma^{alpha} \partial \Sigma^{\beta}} + \Gamma_{ij}^{\gamma}\frac{\partial X^{i}}{\partial \Sigma^{\alpha}}\frac{\partial X^{j}}{\partial \Sigma^{\beta}}\Big]
 \end{equation}\\
 $\Sigma^{\alpha}$ represents the intrinsic co-ordinate parametrizing the shell's surface, where $n_{\gamma}^{\pm}$ denotes the two-sided unit normal for the surface. It takes the form as
 \begin{equation}\label{24}
n_{\gamma}^{\pm} = \pm \Big|g^{ij}\frac{\partial f}{\partial X^{i}}\frac{\partial f}{\partial X^{j}}\Big|^{-\frac{1}{2}}\frac{\partial f}{\partial X^{\gamma}}
\end{equation}\\
where $n^{\gamma}n_{\gamma}=1$. From the Lanczos equation, the surface stress-energy tensor at the interface is given by $S^{q}_{p} = diag(-\Pi,\Xi)$. Here $\Pi$ and $\Xi$ represent the surface energy density and the surface pressure respectively. The following forms determine those parameters as:
\begin{equation}\label{25}
\Pi = -\frac{1}{4\pi R}\Big[\sqrt{e^{-b}}\Big]^{+}_{-}
\end{equation}\\
and\\
\begin{equation}\label{26}
\Xi = -\frac{\Pi}{2} + \frac{1}{16\pi}\Big[\frac{(e^{-b})'}{\sqrt{e^{-b}}}\Big]^{+}_{-}.
\end{equation}\\
Combining equations (\ref{20}) and (\ref{15}) yield the surface density as,
\begin{equation}\label{27}
\Pi = -\frac{1}{4\pi R}\Big[\sqrt{\Lambda R^{2} - \frac{4M}{R}} - \sqrt{He^{-\frac{2\alpha}{3}(\frac{9R^{2}}{2}+\frac{\alpha R^{4}}{4}) + RC_{1}}}\Big]
\end{equation}\\
and then we have the surface pressure as,
\begin{eqnarray}\label{28}
\Xi &=& \frac{1}{48 \pi R^{2}\sqrt{\frac{-4 M}{R} + \Lambda R^{2}}} \times \Big( -12M + 12\Lambda R^{3} + R \sqrt{He^{C_{1}R - \frac{1}{6}\alpha R^{2}(18 + \alpha R^{2})}}
\nonumber \\
&& \sqrt{-\frac{4M}{R} + \Lambda R^{2}} \Big(-6 + 3 C_{1}R + 18\alpha R^{2} + 2\alpha^{2}R^{4}\Big)\Big)
\end{eqnarray}\\
For real solutions we must have $R > \Big(\frac{4M}{\Lambda}\Big)^{\frac{1}{3}}$ and $e^{-\frac{2\alpha}{3}(\frac{9r^{2}}{2}+\frac{\alpha r^{4}}{4}) + rC_{1}} > 0$. The mass of the thin shell can be evaluated using its areal density as,
\begin{eqnarray}\label{29}
M_{Shell} &=& 4\pi R^{2} \pi
\nonumber \\
&&= \Big[\sqrt{R^{2}He^{-\frac{2\alpha}{3}(\frac{9R^{2}}{2}+\frac{\alpha R^{4}}{4}) + RC_{1}}} - \sqrt{\Lambda R^{4} - 4MR}\Big].
\end{eqnarray}\\
The above equation allows us to compute the total mass of the gravastar as a function of the thin shell's mass by,
\begin{eqnarray}\label{30}
M &=& \frac{1}{4R}\times e^{-\frac{2\alpha}{3}(\frac{9R^{2}}{2}+\frac{\alpha R^{4}}{4})}\Big(-R^{2}He^{RC_{1}}-M_{Shell}^{2}e^{\frac{2\alpha}{3}(\frac{9R^{2}}{2}+\frac{\alpha R^{4}}{4})}+
\nonumber\\
&& 2M_{Shell}e^{\frac{2\alpha}{3}(\frac{9R^{2}}{2}+\frac{\alpha R^{4}}{4})}\sqrt{R^{2}He^{RC_{1}-\frac{2\alpha}{3}(\frac{9R^{2}}{2}+\frac{\alpha R^{4}}{4})}}+R^{4}e^{\frac{2\alpha}{3}(\frac{9R^{2}}{2}+\frac{\alpha R^{4}}{4})}\Lambda\Big)
\end{eqnarray}\\

\section{Properties of the gravastar in cylindrical space-time}\label{sec7}

This section has explored various features of the gravastar model within the specific framework of cylindrical space-time.

\subsection{ Proper length of the shell}

The gravastar's interior and the exterior regions are partitioned by the thin shell. We have considered here $r_{1} = D$ and $r_{2} = D + \epsilon$ as the radii of the gravastar's interior and the exterior boundaries respectively. The shell's thickness is assumed to be $\epsilon \ll 1$, implying a small change in the shell length. The thickness here is given by \cite{Mazur},
\begin{equation}\label{31}
l = \int_{D}^{D+\epsilon} \sqrt{e^{b}} dr.
\end{equation}\\
From equation (\ref{18}), we obtain
\begin{equation}\label{32}
l = \frac{1}{\sqrt{8\Lambda}}\Big[\sqrt{\alpha(3 + \alpha r^{2})}\Big(r + \frac{3 \ln(-\sqrt{\alpha}r + \sqrt{3 + \alpha r^{2}})}{\sqrt{\alpha}\sqrt{3 + \alpha r^{2}}}\Big)\Big]_{D}^{D + \epsilon}.
\end{equation}\\
\begin{figure}[h!]
\centering
\includegraphics[scale=0.5]{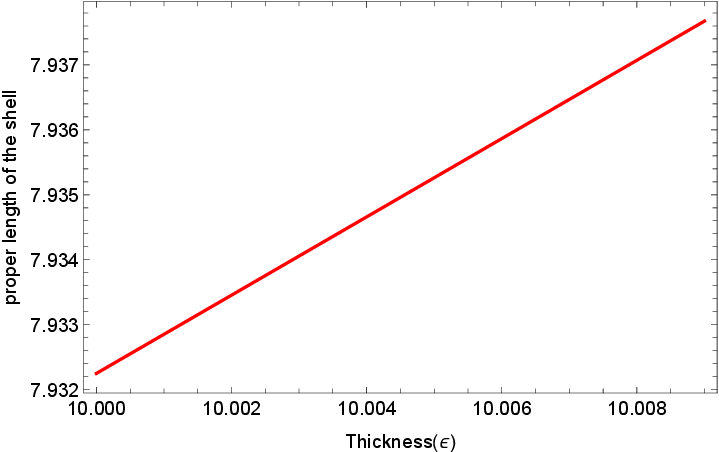}
\caption{The change in the proper length of the shell is plotted with respect to the thickness parameter $\epsilon$(km) $\alpha = 0.01491932683$ \cite{K}}
\end{figure}\\
Figure (\ref{4}) illustrates the variation in the shell's proper length as a function of its thickness. It increases with the shell thickness towards the outer edge of the shell \cite{QT,K,RT}.\\

\subsection{ Energy within the shell}

The EoS $p = -\rho$ in the inner sector implies the presence of negative pressure within it. The emergence of singularity within the gravastar is prevented by the repulsive force generated by the interior region. The shell's thickness plays a crucial role in determining its energy content \cite{GT}. The shell's energy content is expressed as,
\begin{equation}\label{33}
E = 4\pi \int_{D}^{D+\epsilon} \rho r^{2} dr.
\end{equation}\\
From equation (\ref{19}), after substituting $\rho$ we obtain
\begin{eqnarray}\label{34}
E &=& \int_{D}^{D+\epsilon} 4\pi r^{2}A e^{-\alpha r^{2}}dr
\nonumber \\
&&= 4\pi A\Big[\frac{-re^{-\alpha r^{2}}}{2\alpha} + \frac{\sqrt{\pi}erf(\sqrt{\alpha r})}{\alpha^{\frac{3}{2}}}\Big]_{D}^{D+\epsilon}
\end{eqnarray}\\
\begin{figure}[h!]
\centering
\includegraphics[scale=0.5]{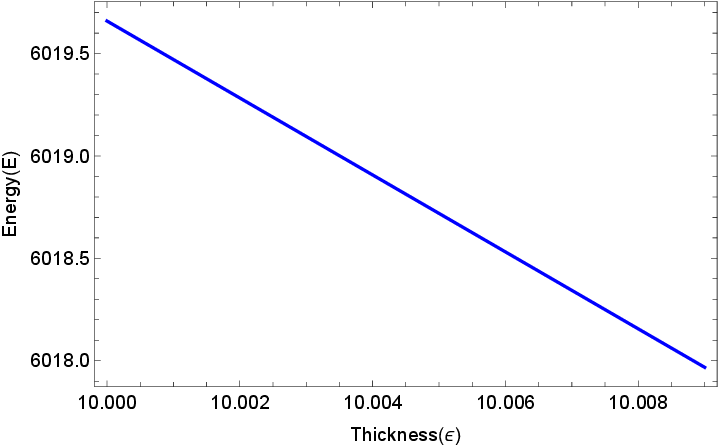}
\caption{The change in the energy content is represented with respect to the thickness parameter $\epsilon$(km) for $\alpha = 0.01491932683$ \cite{K} in the shell region}
\end{figure}\\
here $erf$ denotes the error function. Figure (\ref{5}) illustrates the variation of the energy content$(E)$ with the thickness of the shell. The variation reveals that the shell's energy decreases with the increasing thickness. At the interface between the vacuum-energy filled interior and the shell, which is supported by stiff fluid, energy is relatively high whereas at the shell-exterior boundary, the energy decreases slightly \cite{K}.

\subsection{ Entropy of the shell}

The measure of disorder or disturbance within a celestial entity is referred to as its entropy. The event horizon's size is essential to the process of determining the entropy of the black hole. In contrast, for gravastar the shell thickness affects the entropy, due to the lack of the event horizon. Mazur and Mottola's model proposed that the entropy density within the interior is zero \cite{Mazur,Mottola}. But the shell's entropy is given by \cite{GT},
\begin{equation}\label{35}
S = \int_{D}^{D+\epsilon} 4 \pi r^{2}s(r)\sqrt{e^{b}}dr.
\end{equation}\\
Here $s(r) = \eta \frac{k_{B}}{\hbar}\sqrt{\frac{p}{2\pi}}$ which denotes the entropy density, $\eta$ being a dimensionless parameter. Setting $ k_{B} = \hbar = 1 $ and combining the equations (\ref{18}) and (\ref{19}) yield,
\begin{eqnarray}\label{36}
S &=& \frac{4 \pi \eta}{\sqrt{2 \pi}}\int_{D}^{D+\epsilon} r^{2}\sqrt{A e^{-\alpha r^{2}} \frac{\alpha^{2}r^{2} + 3 \alpha}{2\Lambda}}
\nonumber \\
&&= 2\sqrt{\pi}\eta \sqrt{\frac{A}{\Lambda}}\Big[\frac{\sqrt{e^{-\alpha r^{2}}\alpha r^{4}(3 + \alpha r^{2})}\Big(\sqrt{\alpha}r\sqrt{3 + \alpha r^{2}}(3 + 2\alpha r^{2}) + 18 \tanh^{-1}(\frac{\sqrt{\alpha}r}{\sqrt{3}+\sqrt{3 + \alpha r^{2}}})\Big)}{8\alpha^{\frac{3}{2}}r^{2}\sqrt{3 + \alpha r^{2}}}\Big]_{D}^{D + \epsilon}.
\end{eqnarray}\\
\begin{figure}[h!]
\centering
\includegraphics[scale=0.5]{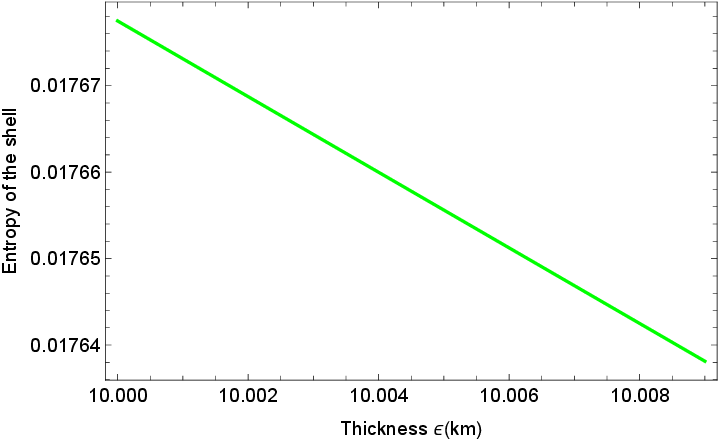}
\caption{The variation in the entropy of the shell is plotted with respect to the thickness parameter $\epsilon$(km) for $\alpha = 0.01491932683$ \cite{K}}
\end{figure}\\
Figure (\ref{6}) illustrates that the entropy decreases steadily as the thickness of the shell grows. As the shell's thickness approaches the outer boundary, the entropy of the gravastar shell also decreases \cite{K}.\\

\section{Stability analysis}\label{sec8}

Analyzing the surface red-shift of gravastars serves as a vital probe for investigating their stability. The formula given by $Z_{s} = \frac{\Delta \lambda}{\lambda_{e}} = \frac{\lambda_{0}}{\lambda_{e}}$ provides a mean to calculate the gravitational red-shift in gravastars. Here the wave-length detected by the observer and the one emitted from the source are denoted by $\lambda_{0}$ and $\lambda_{e}$ respectively. It was proposed that the surface red-shift of an isotropic, stable and perfect fluid configuration should not exceed 2 \cite{Bu,Str}. For anisotropic fluids, the surface red-shift was suggested to possibly reach as high as 3.84 \cite{Iv}. Furthermore, research indicated that for isotropic fluids the condition $Z_{s}\leq 2$ holds true when there is no cosmological constant \cite{DE}. It was also found that for anisotropic stars, the surface red-shift $Z_{s}\leq 5$, considering the effects of the cosmological constant \cite{CG}. We have determined the surface red-shift using the following mathematical expression as \cite{QT,K,Ban}\\
\begin{equation}\label{37}
Z_{s} = -1 + \mid g_{tt} \mid ^{\frac{-1}{2}} = \frac{1}{\sqrt{e^{\alpha r^{2} + 2 \ln \beta}}} -1.\\
\end{equation}\\
\begin{figure}[h!]
\centering
\includegraphics[scale=0.5]{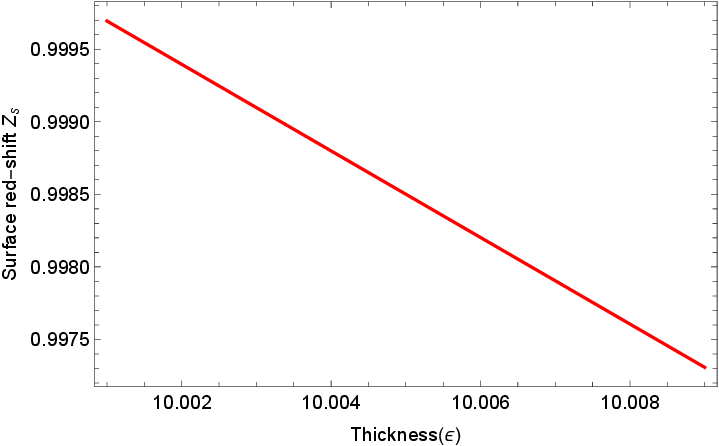}
\caption{The variation of $Z_{s}$ is plotted with respect to the thickness parameter $\epsilon$(km) for $\alpha = 0.01491932683$ and $\beta = 0.2371387430$ \cite{K}}
\end{figure}\\
Figure (\ref{7}) provides a graphical illustration of the surface red-shift $Z_{s}$. The figure reveals that the value of $Z_{s}$ stays within the stability zone throughout the shell. Consequently, the present gravastar model emerges as a stable and acceptable solution.\\

\section{Conclusion and discussions}\label{sec9}

Building on the Mazur-Mottola model from  a General Relativistic perspective, we have designed a distinctive gravastar stellar model in cylindrical space-time, utilizing the Kuchowicz metric potential. Our gravastar model consists of three distinct regions: the interior, the thin shell and the exterior each having a different EoS. The interior region is composed entirely of dark energy \cite{Mazur,Mottola}. The following are the key characteristics of gravastars. In the interior region we have obtained a non-singular metric function, as expressed in equation (\ref{15}). This metric potential is finite, remains positive and is well-defined throughout the entire interior region as is evident from figure (\ref{1}). Thus our gravastar model in cylindrical space-time successfully eliminates the central singularity typically associated with the black holes.\\
Using the thin-shell approximation, we have calculated the metric potential in the shell region as given by equation (\ref{18}), which stays finite and positive throughout its entirety as shown in figure (\ref{2}). Additionally by applying energy-conservation principles, we were able to derive an expression for the pressure and matter inside the shell. Figure (\ref{3}) graphically presents the variation(or matter density) as a function of the thickness parameter ($\epsilon$). It shows that the matter density inside the shell decreases steadily as one moves towards the exterior. The shell consists of an ultra-relativistic fluid, given that the pressure or the matter density decreases monotonically towards the outer surface, physically it indicates that the amount of stiff matter is declining towards the outer edge of the shell. For the thin shell formation, we have considered the junction conditions connecting the interior and the exterior space-times. Using the Darmois-Israel junction condition we have derived the surface energy density and the surface pressure there. In addition, we have obtained the maximum allowed value for the radius, and have calculated the mass of the shell based on its surface energy density.\\

Physical features of the model: By examining the geometric characteristics of the intermediate thin shell, we have investigated various physical properties associated with it.
\begin{itemize}
\item {Proper length}: Figure (\ref{4}) illustrates how the proper length changes as a function of the thickness parameter($\epsilon$). The figure shows that the proper length exhibits a uniform increasing nature with growing shell thickness. The monotonic increase in proper length of the gravastar is consistent with the research findings \cite{QT,S,SS}.
\item {Energy}: Figure (\ref{5}) depicts the variation in the shell energy. It shows that the energy content within the shell decreases with growing thickness. The energy fluctuations behaves similarly to the variations in the matter density.
\item {Entropy}: Figure (\ref{6}) gives an illustration of the shell entropy's behavior, revealing a decline in the entropy as the shell thickness increases \cite{K}.\\
\end{itemize}

At last, surface red-shift analysis was performed to examine the stability of the model in our study. It was well within the stability range. Thus our research shows that the proposed model exhibits stability in the context of cylindrical space-time. In conclusion, this study presents an alternative perspective on gravastars, utilizing the Kuchowicz metric potential in cylindrical space-time. Considering s single parameter complicates the process of determining physically viable solutions for the remaining parameters. However, our investigation reveals that all the obtained solutions are physically valid, finite and well-behaved at the origin. Thus this gravastar model based on the Kuchowicz metric potential in cylindrical space-time appears to be theoretically consistent and physically plausible. These findings motivate to explore further research opportunities using various other metric potentials. Although gravastars remain experimentally unobserved, the theoretical discussions in the literature suggest potential connections to gravitational wave detection by LIGO, whether the detected gravitational waves(GW) originate from black holes or gravastars merging. This study demonstrates the theoretical feasibility and physical viability of gravastar existence.\\

\end{document}